\documentclass[12pt]{iopart}
\expandafter\let\csname equation*\endcsname\relax
\expandafter\let\csname endequation*\endcsname\relax
\usepackage{amssymb}
\usepackage{amsmath}
\usepackage{amsthm}
\usepackage{amsfonts} 
\usepackage{bm}
\usepackage{physics}
\usepackage{graphics}
\usepackage{braket} 
\usepackage{graphicx}
\usepackage{dcolumn}

\begin{document}

\title[]{Calculating the Coulomb blockade phase diagram in the strong coupling regime of single-electron transistor: a quantum Monte Carlo study}

\author{Pipat Harata, Wipada Hongthong and Prathan Srivilai*}

\address{NanoMaterials Physics Research Unit (NMPRU), Department of Physics,
Faculty of Science, Mahasarakham University, Khamriang Sub-District, Kantarawichai District, Mahasarakham 44150, Thailand}
\ead{prathan.s@msu.ac.th}
\vspace{10pt}
\begin{indented}
\item[]January 2024
\end{indented}

\begin{abstract}
We present a novel approach for calculating the Coulomb Blockade Phase Diagram (CBPD) in the experimentally accessible strong coupling regime of a single-electron transistor (SET). Our method utilizes the Path Integral Monte Carlo (PIMC) technique to accurately compute the Coulomb oscillation of the Differential Capacitance (DC). Furthermore, we investigate the impact of the gate voltage and temperature variations on the DC, thereby gaining insights into the system's behaviour. As a result, we propose a method to calculate the Coulomb Blockade Boundary Line (CBBL) and demonstrate its efficacy by setting the visibility parameter to $10\%$. The resulting boundary line effectively defines the transition between the Coulomb and non-Coulomb blockade regimes, thereby enabling the construction of a comprehensive CBPD.
\end{abstract}

%
\vspace{2pc}

\section{Introduction}
\label{sec:intro}

Single-electron devices \cite{ GrabertDevoret, Likharev1999} have attracted attention continuously due to their nanoscale size, low power dissipation, and ability to offer novel functionalities. However, realizing the Coulomb blockade effect in nanoscale junction systems necessitates fulfilling two crucial conditions \cite{GrabertDevoret, Devoret1992}. Firstly, the kinetic energy of the electrons must be considerably lower than the charging energy to suppress thermal fluctuations that could obscure the quantization effect. Secondly, electrons must be confined or localized on the island, necessitating a tunnelling barrier with resistance much more significant than the quantum resistance unit. These conditions are essential to ensure the proper operation and manifestation of the Coulomb blockade effect. The focus of the theoretical study is on the optimization of these two critical Coulomb blockade effect parameters. This optimization is represented as a CBPD, which features a boundary line separating the Coulomb blockade regime from the Non-Coulomb blockade regime \cite{Wang2014}. By referring to the CBPD, researchers can determine the extent of the Coulomb blockade effect in investigations involving single-electron devices. 

A low-conductance SET can be defined as its dimensionless parallel conductance, $g < 1 $, where $g=(G_{S}+G_{D})/G_{K}$, with $G_{S}$ and $G_{D}$ are the high-temperature conductance of tunneling junctions connected with source and drain electrodes, as depicted in Fig.\,\ref{fig:SET}(a), and the quantum conductance $G_{K}=e^2/h$. Applying Green's non-equilibrium function approach \cite{Damle2001, Taylor2001, Louis2003}, the CBPD for a low-conductance SET was calculated \cite{Wang2014} and referred to as a weak coupling regime \cite{Grabert1994, Konig1997, Konig1998}. Utilizing the proposed method, one can determine the optimum temperature to operate a low-conductance SET for various applications, such as single-electron memory \cite{Amine2013, Kouta2019, Nakazato1994} and quantum dots \cite{Choi1998}, which trap and manipulate individual electrons, allowing researchers to explore quantum phenomena. On the other hand, a high-conductance SET, i.e., $g \geq 1$, is well-known for its sensitivity to single-electron charges, which qualifies it for terahertz electronics \cite{Asgari2021, Park2004}, ultra-sensitive sensor applications \cite{Nakajima2016, Cleland1993}, and qubits in quantum computing \cite{Aassime2001}. The CBPD, however, cannot be determined using Green's non-equilibrium function method since it exceeds the limits of perturbation theory \cite{Grabert1994, Konig1997, Konig1998}. As a result, there needs to be more theoretical study in the field of nanoscience. Consequently, this paper proposes a calculation of the CBPD for a high-conductance SET, known as a strong coupling regime \cite{ Schoeller1994, Joyez1997, Wang1997, Herrero1999, Goppert2000}.  

This paper presents an approach for calculating the CBPD within the experimentally accessible strong coupling regime \cite{ Wallisser2002}. Our method aims to characterize the charge fluctuation within the SET. Initially, we calculate the Coulomb oscillation of the DC \cite{Jezouin2016, Idrisov2017} by the PIMC technique \cite{Ceperley1995, Negele1987, Troyer2005}, which provides valuable insights into the system's behavior. Subsequently, we calculate the visibility parameter \cite{ Jezouin2016, Idrisov2017, Limmer2013}, quantifying the disparity between the minimum and maximum magnitudes of the DC. By leveraging the visibility parameter, we can calculate the boundary line separating the Coulomb blockade regime from the Non-Coulomb blockade regime. This proposed methodology allows for a more thorough investigation of the behavior of the SET system and provides an enhanced understanding of the system's behavior without coupling limitations.    

The sections of this paper are structured as follows: Section 2 provides an introduction to the Hamiltonian governing the SET system, along with basic notations pertinent to the path integral representation employed for the DC calculations. Section 3 presents the results of the DC simulation by utilizing the PIMC technique. Additionally, this section explores the impact of the temperature variations and tunnelling conductance on the measured DC. Subsequently, Section 4 introduces the calculation for the visibility parameter and determines the precise boundary line and the CBPD. Lastly, in Section 5, we offer concluding remarks on the findings of this study and discuss potential avenues for future research in this domain.

\section{Model Hamiltonian and Differential Capacitance Calculation}
\label{sec:model} 
\subsection{Model Hamiltonian}
\begin{figure}
\centering
\includegraphics[width=0.90\textwidth]{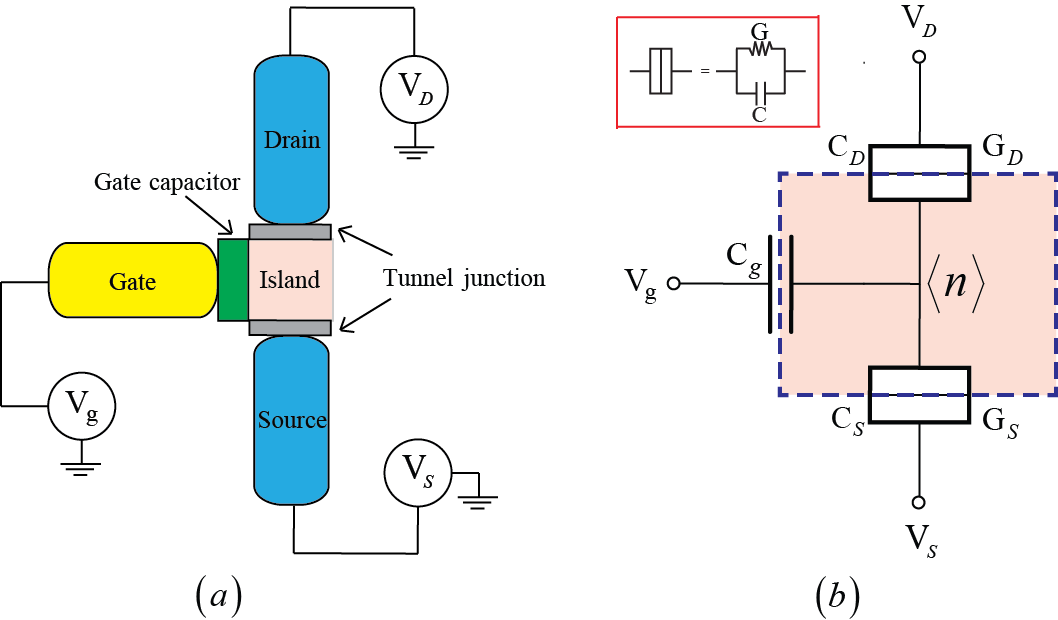}
\vspace{0.2cm}
\caption{ According to the schematic structure (a), the SET system comprises two tunnel junctions and a capacitive junction. The voltage differential,$V_{D}-V_{S}$, defines the bias configuration, while the gate voltage $V_{ g}$  is directly coupled to the island via the gate capacitance $C_{g}$. The central portion of the SET, denoted by the blue dashed line, encompasses the region where excess electrons $n$ are contained (b)} 
\label{fig:SET}
\end{figure}
Consider the arrangement in Fig.\,\ref{fig:SET}(a), where the voltage gate controls the electrostatic potential on a metallic island connecting two tunnel junctions, and Fig.\,\ref{fig:SET}(b) shows its equivalent circuit. In this two primary dimensionless parameters can govern the transport properties of the SET \cite{Wallisser2002}. Firstly, the dimensionless inverse temperature  $\beta E_{C}$ is the ratio of the charging energy to the electron’s kinetic energy, where the charging energy $E_{C}=e^{2}/2C_{\Sigma}$ with the total capacitance $C_{\Sigma}=C_{S}+C_{D}+C_{g}$ and $\beta =1/k_{B} T$. Secondly, the dimensionless parallel conductance, $(g)$, represents the strength of tunneling phenomena, accessible by measuring the high-temperature conductance of two tunneling junctions \cite{Wallisser2002}. 

The quantum transport properties of the SET can be effectively described by considering the electrostatic charge and electron tunnelling between the leads and the island. Furthermore, in the metallic island system, the impact of energy quantization can be disregarded due to the negligible spacing between the island states on the metallic island compared to the charging energy. Consequently, it is possible to construct a microscopic model of the SET using the Hamiltonian represented by \cite{GrabertDevoret} 

\begin{equation} \label{Hamiltonian}
H=H_{B}+H_{T} + H_{C}\, ,
\end{equation}
where the conduction electrons of the leads and the island are expressed by
\begin{equation}
\label{HsubB} H_{B}=\sum_{Jk\sigma}\epsilon_{Jk\sigma}c_{Jk\sigma}^{\dag}c_{Jk\sigma}^{}+
\sum_{k\sigma}\epsilon_{k\sigma}d_{k\sigma}^{\dag}d_{k\sigma}\, ,
\end{equation}
respectively. Here $\epsilon_{Jk\sigma}$ is the energy of an electron with a longitudinal wave vector $k$ in channel $\sigma$ of lead $J$, where $J \in \{S, D\}$. Likewise, $\epsilon_{k\sigma}$ is the energy of an electron on the island. $c_{Jk\sigma}$ and $d_{k\sigma}$ are the corresponding electron annihilation operator of lead $J$ and the island with longitudinal wave vector $k$ in channel $\sigma$.

The tunnelling of the electrons across the two tunnel junctions is described by
\begin{eqnarray}
\label{Htunneling}
H_{T}=\sum_{kq\sigma}&\Big[&d_{k\sigma}^{\dag}t_{Skq \sigma}{\rm e}^{-i\varphi} c_{Sq\sigma}
+c_{Dk\sigma}^{\dag}t_{Dkq \sigma}{\rm e}^{i\varphi} d_{q\sigma}+\hbox{H.c.}\label{HsubT}\Big],
\end{eqnarray}
where $t_{Skq\sigma}$ is the amplitude for an electron in state $\ket {q \sigma}$ on the source lead tunnel onto the island with the final state $\ket {k \sigma}$. Likewise, $t_{Dkq \sigma}$ denotes the electron tunnelling amplitude of the tunnel junction between the island and the drain lead. The charge shift operator ${\rm e}^{-i\varphi}$ adds one charge to the island, corresponding with the phase operator $\varphi$ conjugating to the number operator $n$ of the excess charge on the island. Here we defined $\varphi_{S}=\varphi_{D}= 0$ and $ \hbar=1$ through this letter. 

The charging energy, denoted as $H_{C}={{E_{C}}{( {n}-{n_{g}})}^{2}}$, represents the effect of the (excess) electron numbers $n$ on the island, and $n_{g} = C_{g}V_{g}/e$ is the continuous charge induced by the gate voltage. By starting from the Hamiltonian presented in Eq.(\ref{Hamiltonian}), the evaluation of the grand partition function,  $Z={\rm tr}\, \{{\rm e}^{-\beta (H-\mu\hat{N})}\}$, for the SET can be accomplished by applying the imaginary-time path integral representation \cite{Negele1987} and utilizing the large channel approximation \cite{ Grabert1994, Ambegaokar1982, Goppert1999}. In this context, $\mu$ and $\hat{N}$ stand for the system's chemical potential and the particle number operator, respectively. However, we skip the detailed calculation procedure as it has been thoroughly addressed in the provided reference \cite{Christoph2004}. Therefore, all possible paths are expressed in imaginary time, and we can perform the quantum Monte Carlo method \cite{Ceperley1995, Negele1987, Troyer2005} to analyse the quantum statistical characteristics of the SET. 

\subsection{Differential Capacitance Calculation}

 The average excess tunnelling charge can effectively characterize the quantum transport properties and Coulomb blockade effect of the SET \cite{ Konig1997, Joyez1997, Goppert2000, Konig1998_1, Chouvaev1999, Devoret2000}. In particular, the variation in the average tunnelling charge on the island concerning the gate voltage can be represented by $C_{\rm{diff}} = \partial {\langle Q_{t} \rangle}/ \partial {V_{g}}$ \cite{Jezouin2016, Idrisov2017}. Here the average tunnelling charge is $\langle Q_{t}\rangle= -e\langle n \rangle $, and $\langle n \rangle$ represents the island's average (excess) electron number. Consequently, the average electron number can be calculated by $\left\langle n \right\rangle  = {n_g} + (2\beta {E_C})^{-1}({\partial( \ln Z)}/{\partial {n_g}})$ \cite{Goppert2001}. As a result, the DC of the SET can be expressed in terms of the winding number and represented by
\begin{equation}\label{cdiff}
    \frac{C_{\rm{diff}}}{C_{g}} = \frac{{d\left\langle n \right\rangle }}{{d{n_g}}}=1 - \frac{{2{\pi ^2}}}{{\beta {E_C}}}\Big(\left\langle {{k^2}} \right\rangle -\left\langle {{k}} \right\rangle^{2}\Big),
\end{equation}
where the winding number expectation value can be represented in terms of the path integral representation as 

\begin{equation}\label{expect}
\left\langle {{X}} \right\rangle  = \frac{\sum\limits_{k=-\infty}^{\infty} {\int\limits_{\xi \left( 0 \right)=0}^{\xi \left( {\beta {E_C}} \right) =0} {D\xi \;{X}{e^{ - S[\xi ,k]}}} }}{\sum\limits_{k=-\infty}^{\infty} {\int\limits_{\xi \left( 0 \right) = 0}^{\xi \left( {\beta {E_C}} \right)=0} {D\xi \,{e^{ - S\left[ {\xi ,k} \right]}}} }}.  
\end{equation}
Here all energies are measured in units of the charging energy $E_{C}$ to make the PIMC simulation more convenient. The partition function, typically expressed as a path integral over the phase variable $\varphi$, has been obtained through the transformation $\varphi (\tau)=\xi(\tau) +\nu_{k}\tau$ with $\nu_{k}=(2\pi k/\beta {E_C})$, and the utilization of the boundary condition of the paths ${\xi}\left( 0 \right) = {\xi}\left( {\beta {E_C}} \right)$. As a result, the Euclidean action of the SET, denoted as $S[\xi, k]=S_{C}[\xi, k]+S_{T}[\xi, k]$, can be represented in terms of the Coulomb action,
\begin{equation}\label{CS}
S_C\left[ \xi, k \right]\, = \,\int\limits_0^{\beta {E_C}} {d\tau \left( {\frac{{\mathop {{\dot{\xi} ^2}}}}{4}} \right)} +\frac{4\pi^{2} k^{2}}{\beta E_{C}} + 2\pi i k n_{g},
\end{equation}

where $\dot{\xi} = (d\xi /d \tau) $ and the tunnelling action, 
\begin{equation}\label{TS}
S_T \left[ \xi, k \right]\, = \, - g\int\limits_{0}^{\,\beta {E_C}} d \tau \,\int\limits_{0}^{\,\beta {E_C}} d {\tau^{\prime}}\,\alpha\left(\tau-\tau^{\prime}\right)\,\cos \left( {\xi(\tau) -\xi ( \tau^{\prime})} +\nu_{k} (\tau-\tau^{\prime})\right),
\end{equation}
where the tunnelling kernel reads 
 \begin{equation}
\alpha \left( \tau  \right) = \frac{1}{{4{{\left( {\beta {E_C}} \right)}^2}{{\sin }^2}\left( {\frac{\pi }{{\beta {E_C}}}\tau } \right)}}.
\end{equation}
The exact solution for the expectation value of the winding number, as described in Eq.(\ref{expect}), cannot be obtained through analytical calculations due to its non-Gaussian integral nature. However, the winding number's expectation value and the DC of the SET can be evaluated using the PIMC technique. 

\subsection{ PIMC simulation of the winding number expectation value}

In this subsection, we summarized the essential details of the PIMC simulation for the SET following the main idea in Ref.\,\cite{ Christoph2004}. The PIMC technique was employed to simulate the DC results of the SET across the entire range of the dimensionless gate voltage, $ n_{g}\in \{ 0,1\}$, the inverse temperature, $\beta E_{C}\in \{ 1,21\}$, and specifically at high dimensionless conductance, $ g\in \{ 1,15\}$, where the perturbation theory fails. In order to utilize the Metropolis algorithm \cite{Metropolis2004, Grotendorst2002}, a positive definite action is required. Consequently, we must include the imaginary part of the Coulomb action in the observable to be measured, i.e., $\exp{-2 \pi i n_g k}$, and use the fact that the expectation value of the odd function is zero.  The expectation values in Eq.(\ref{cdiff}) can be rewritten as 
\begin{equation}\label{avgk}
\left\langle {{k}} \right\rangle = \frac{\left\langle k \sin{(2 \pi n_g k )}
\right\rangle_0}{\left\langle \cos{(2 \pi n_g k)} \right\rangle_0},
\end{equation}
and 
\begin{equation}\label{avgk2}
\left\langle {{k^2}} \right\rangle = \frac{\left\langle k^{2} \cos{(2 \pi n_g k )}
\right\rangle_0}{\left\langle \cos{(2 \pi n_g k)} \right\rangle_0},
\end{equation}
where $\left\langle X \right\rangle_0$ denotes the expectation value with the positive action
\begin{equation}
\label{eq:s0_disc} S_{0}[\xi,k] = \,\int\limits_0^{\beta {E_C}} {d\tau \left( {\frac{{\mathop {{\dot{\xi} ^2}}}}{4}} \right)} +\frac{4\pi^{2} k^{2}}{\beta E_{C}} - S_T \left[ \xi, k \right]. 
\end{equation}
The discrete formula of Eq.(\ref{avgk}) used in the PIMC simulation is
\begin{equation}
\left\langle {{k}} \right\rangle = \frac{ \sum_{k_{j}\in \{k,\xi\}} k_{j} \sin\left( 2 \pi k_{j} n_g
\right)}{\sum_{k_{j}\in\{ k,\xi\}} \cos\left( 2 \pi k_{j} n_g \right)},
\end{equation}
where the sum $\sum_{k_{i}\in\{ k,\xi\}}$ stands for the accumulation of $k_j$ over the configurations $\xi$ and $k$ sampled by the Metropolis algorithm  for the positive action in Eq.(\ref{eq:s0_disc}). The expectation value in Eq.(\ref{avgk2}) can proceed similarly. The results for the winding number expectation values over the whole range of gate voltages can obtained from a single Monte Carlo simulation since the positive action is independent of $n_g$. In addition, the DC results were obtained through extensive Monte Carlo simulations, employing multiple measurements to ensure the statistical error remained below a certain threshold, i.e., $ 1\%$, across the entire parameter ranges. Consequently, the error bars displayed in the Monte Carlo data indicate that the magnitude of one standard deviation is smaller than the symbol size depicted in all the following figures. 
\begin{figure}
\centering
\includegraphics[width=0.85\textwidth]{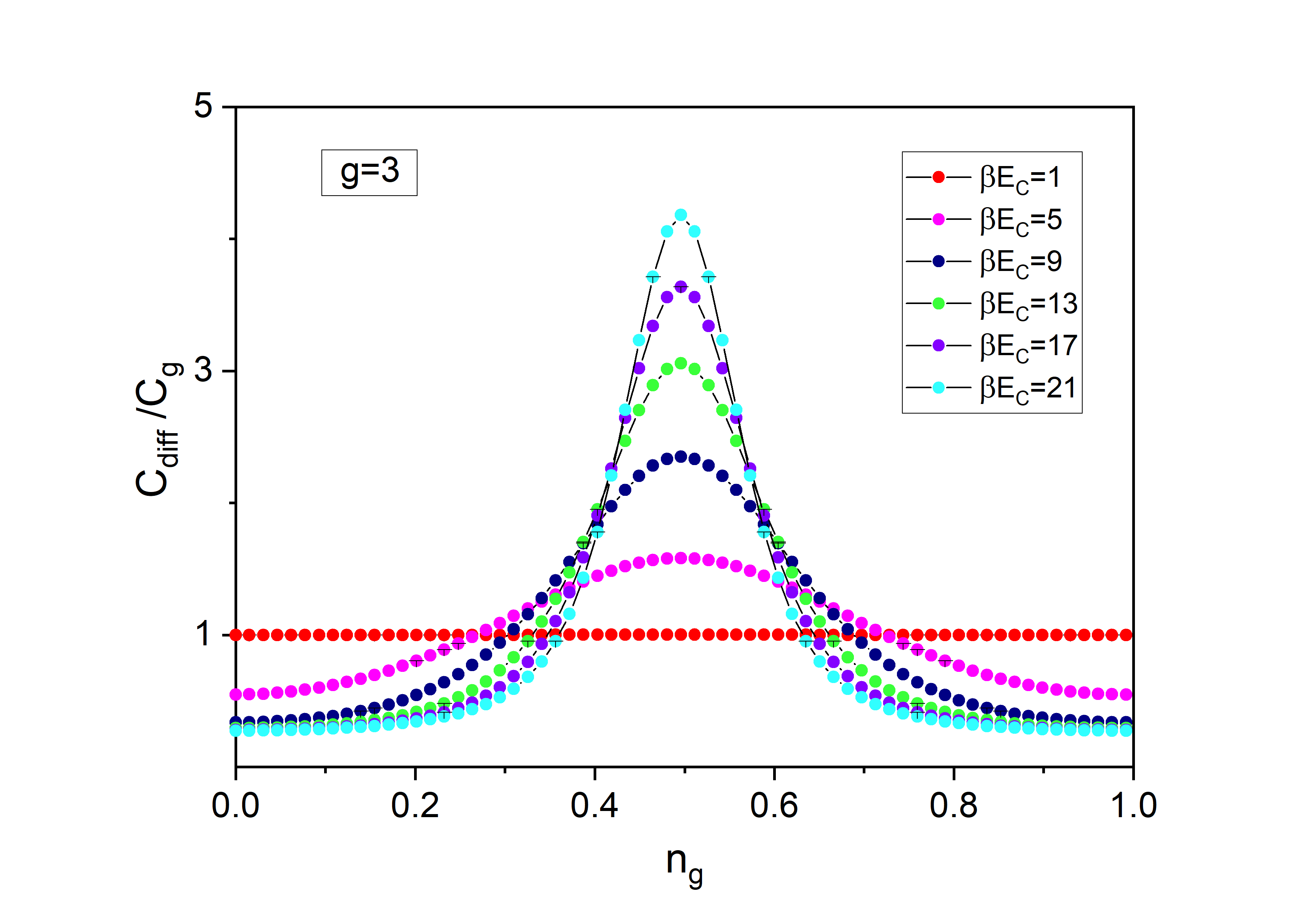}
\vspace{0.2cm}
\caption{ DC of SET exhibits Coulomb oscillations as a function of the dimensionless gate voltage $n_{g}$, where the dimensionless conductance is fixed at $g=3$, while the inverse temperature varies.} 
\label{fig:SET_beta}
\end{figure}

\begin{figure}
\centering
\includegraphics[width=0.85\textwidth]{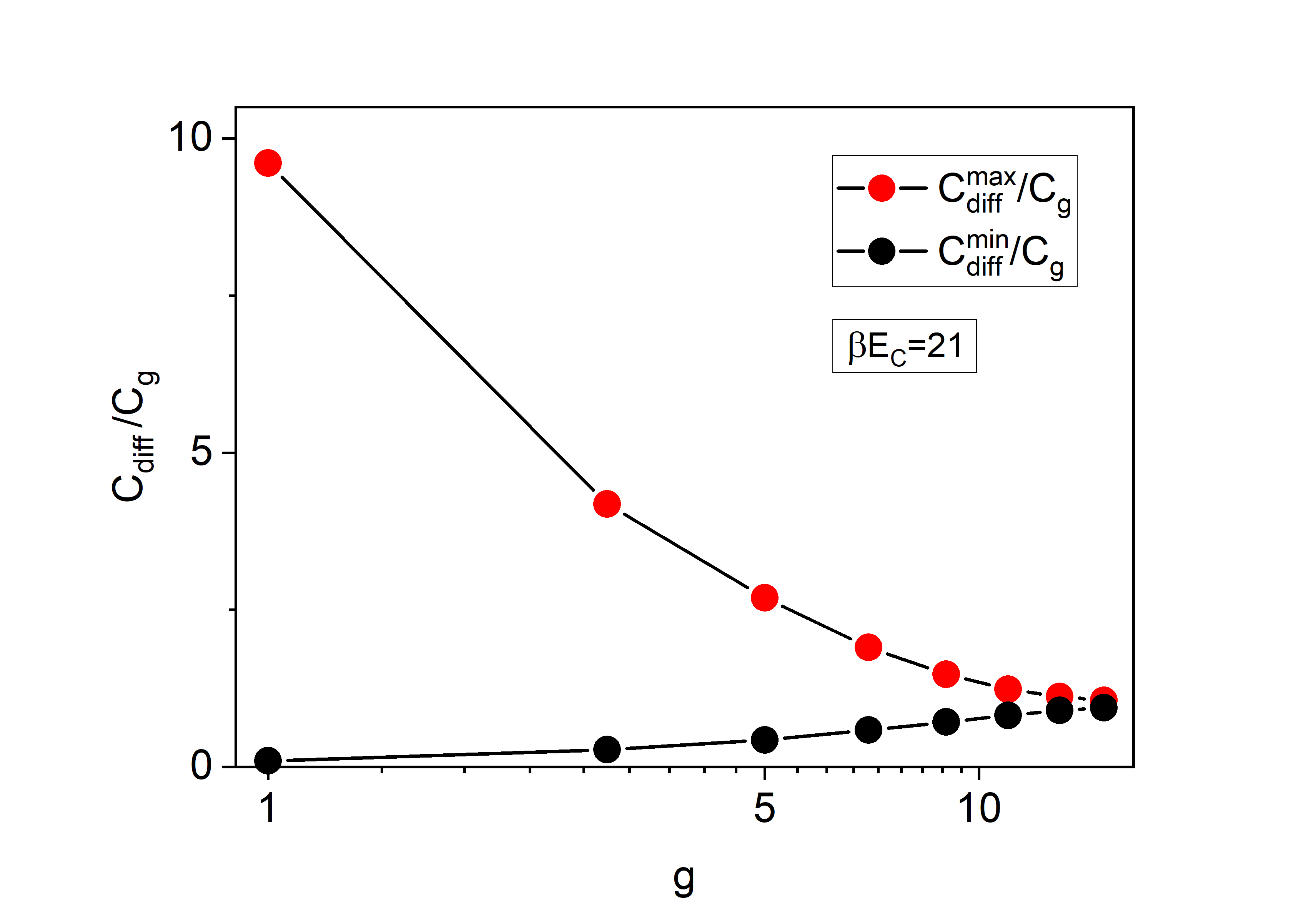}
\vspace{0.2cm}
\caption{ SET's normalized maximum and minimum DC obtained from PIMC simulations, plotted against the inverse temperature $(\beta E_{C}=21)$ and normalized to the gate capacitance.} 
\label{fig:Cgg}
\end{figure}

\section{Results and Discussions}

The DC results of the SET were simulated using the PIMC technique, focusing on the dimensionless gate voltage range denoted as $n_{g} \in \{0,1\}$. Fig.\,\ref{fig:SET_beta} presents the results for two specific conditions: the absence of any bias between the source and drain electrodes and the strong tunnelling regime characterized by $g \geq 1$. In Fig.\,\ref{fig:SET_beta}, particular attention is given to the red dotted line, revealing an interesting observation, in which, when $\beta E_{C}=1$, the ratio of the DC to the gate capacitance remains constant at unity, independent of the dimensionless gate voltage. This finding implies that the Coulomb blockade effect ceases to exist since the charging energy becomes approximately equal to the kinetic energy of the electrons. Conversely, as the temperature decreases with a corresponding increase in $\beta E_{C}$, the DC depends on the dimensionless gate voltage. Notably, the magnitudes of the DC peaks reach their maximum values when $n_{g}=0.5$ and their minimum values when $n_{g}=0.0$, as illustrated in Fig\,\ref{fig:SET_beta}. Consequently, it becomes evident from the presence of these peaks in the DC that the Coulomb blockade effect is indeed occurring, provided that the condition $\beta E_{C} \gg 1$ is satisfied. As described above, the Coulomb oscillation peaks provide clear evidence of the occurrence of the Coulomb blockade effect. The electron localization on the island can be governed by the optimization between parameters $ g $ and $\beta E_{C} $, leading to the observed oscillatory behavior in the DC.

\begin{figure}
\centering
\includegraphics[width=0.85\textwidth]{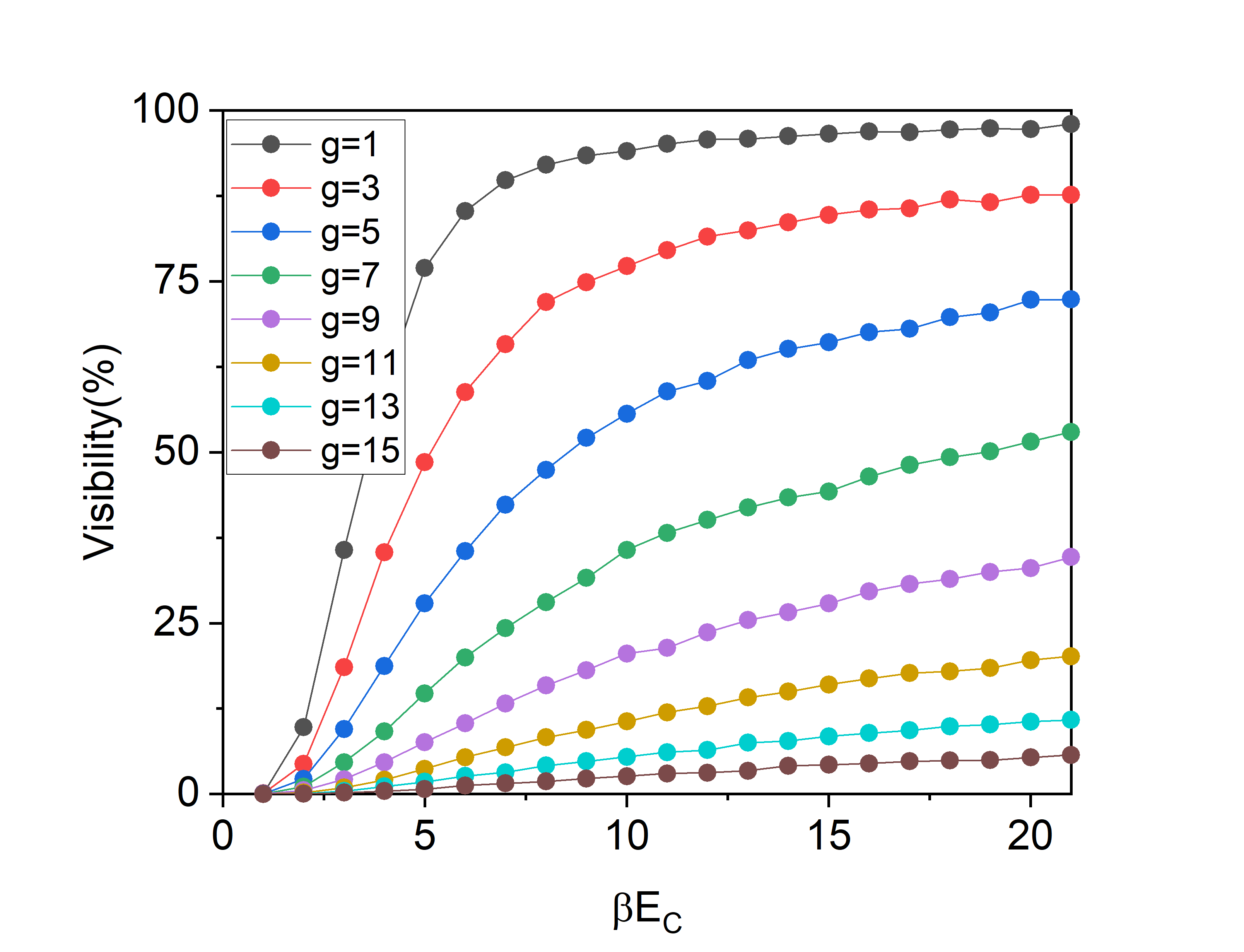}
\vspace{0.2cm}
\caption{ Visibility of SET varies as a function of the inverse temperature for different dimensionless conductance values. } 
\label{fig:visibility}
\end{figure}

Due to the Coulomb blacked effect in the SET being reflected by the Coulomb oscillation in the DC, we can define the minimal and maximal DC at $n_{g}=0.0$ and $n_{g}=0.5$, denoted by $C_{\rm{diff}}^{\rm{min}}$ and $C_{\rm{diff}}^{\rm{max}}$, respectively. By fixed temperature, while the dimensionless conductance varies, the Coulomb oscillation of the DC occurs, reflected as the difference between $C_{\rm{diff}}^{\rm{min}}$ and 
$C_{\rm{diff}}^{\rm{max}}$, as shown in Fig.\,\ref{fig:Cgg}. When $g$ increases significantly, the $C_{\rm{diff}}^{\rm{min}}$ merges with $C_{\rm{diff}}^{\rm{max}}$, and the Coulomb blockade effect vanishes. As a result, it is indicated that the difference in the magnitudes of $C_{\rm{diff}}^{\rm{min}}$  and  $C_{\rm{diff}}^{\rm{max}}$ can identify the Coulomb Blockade and Non-Coulomb Blockade regimes. Furthermore, the magnitude of $C_{\rm{diff}}^{\rm{min}}$ is directly associated with the effective charging energy defined as $ E_{C}^{\ast}/E_{C} = 1-C_{\rm{diff}}^{\rm{min}}$. The effective charging energy is widely recognized as a measure of the strength of the Coulomb blockade effect and this has been extensively investigated in the literature \cite{ Herrero1999, Werner2005, Werner2006}.

\begin{figure}
\centering
\includegraphics[width=0.85\textwidth]{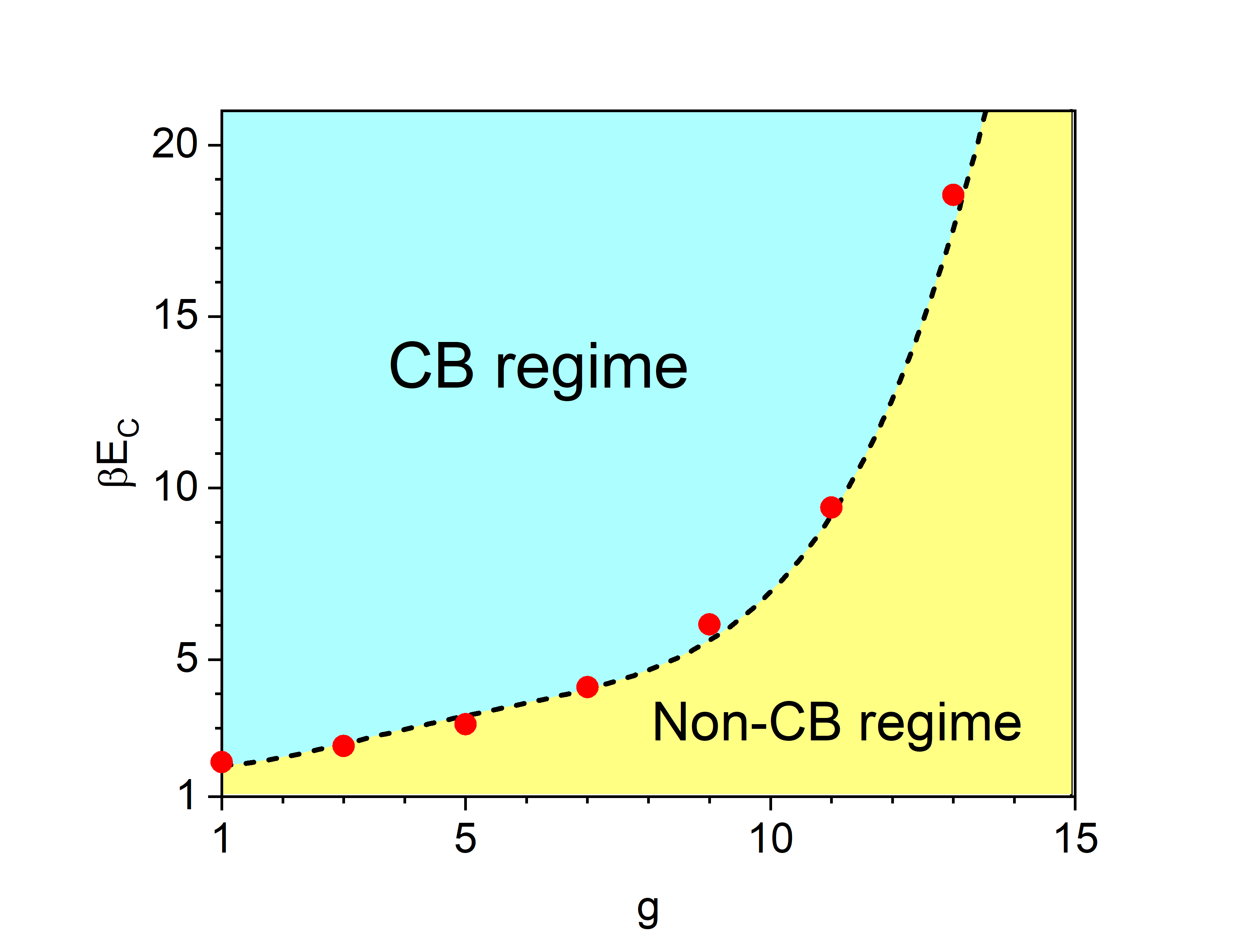}
\vspace{0.2cm}
\caption{ CBPD of SET was calculated with the visibility of $10\% $, illustrating the boundary line that demarcates the Coulomb and Non-Coulomb blockade regimes. The red circle data points correspond to the results of fitting the PIMC data.} 
\label{fig:CBL}
\end{figure}

\subsection*{Visibility}

More practically, to propose a calculation for a CBBL, the visibility of the SET was defined as
\begin{equation}\label{visibility}
    \mathbf{V} = \frac{ C_{\rm{diff}}^{\rm{max}}-C_{\rm{diff}}^{\rm{min}} }{C_{\rm{diff}}^{\rm{max}}+ C_{\rm{diff}}^{\rm{min}} }.
\end{equation}
The inverse temperature's effect on the SET's visibility is illustrated in Fig.\,\ref{fig:visibility}.  Consequently, as $\beta E_{C}$ increases, the visibility also increases and stabilizes at a constant value determined by the $g$ parameter. Therefore, a CBBL is established through the utilization of Monte Carlo data fitting for the individual parameter $g$, employing a polynomial function, $\mathbf{V} = C_{0}+C_{1}(\beta E_{C})+C_{2}(\beta E_{C})^2$, characterized by constants $C_{0}, C_{0}, \rm {and}, C_{2}$. By assigning a desired visibility value $\mathbf{V}$, the polynomial equation can be solved to determine the corresponding $\beta E_{C}$ value. Consequently, constructing the CBBL involves utilizing a set of ordered pairs $(g,\beta E_{C})$. 

As a demonstration, Fig.\,\ref{fig:CBL} illustrates the CBBL within the phase diagram, providing insights into the SET behaviour. The visibility is set explicitly to $\mathbf{V}=10\%$, corresponding to $ C_{\rm{diff}}^{\rm{min}}/ C_{\rm{diff}}^{\rm{max}} \sim 0.82$. Mathematically, the CBBL can be described by the polynomial equation as 
\begin{equation}\label{CBL}
   \beta E_{C} = 2.6465-0.9896g+0.45g^{2}-0.0603g^{3}+0.003g^{4}.
\end{equation}
When the condition $\beta E_{C} > 2.6465-0.9896g+0.45g^{2}-0.0603g^{3}+0.003g^{4}$ is satisfied, the SET operates within the Coulomb blockade region, denoted by the sky blue area. Conversely, if $\beta E_{C} < 2.6465-0.9896g+0.45g^{2}-0.0603g^{3}+0.003g^{4}$, the SET operates outside the Coulomb blockade regime, represented by the yellow region. 

Furthermore, the CBBL represented by Eq.(\ref{CBL}) can be verified with the SET experiments, known for their parallel dimensionless conductance and temperature operation in terms of $\beta E_{C} $. For example, the experiment of the SET with the dimensionless conductance of $ g = 4.75 $ was reported as the Coulomb oscillations of conductance \cite{ Wallisser2002, Christoph2004}. It is easy to predict from Eq.(\ref{CBL}) that the Coulomb blockade effect occurs by operating with $\beta E_{C} > 3.16$. The finding agrees with the Coulomb oscillation experiment shown in Fig.\,6.12 of Ref.\,\cite{Christoph2004}, wherein the Coulomb oscillation peak appears evidently with $\beta E_{C} > 3.0$. Let us next focus on the first experiment demonstrating a SET operating at room temperature \cite{Pashkin2000}. The aluminum SET fabricated by standard e-beam lithography consists of $ g = 0.002 $ and was operated at $ 300 \,K $. From Eq.(\ref{CBL}), one obtains that the condition of the Coulomb blockade occurring is $\beta E_{C} > 2.644$, corresponding with $ 504.28 \,K$, where $E_{C}= 115 \,meV $. The result confirms that the SET can exhibit the Coulomb blockade effect at room temperature. 

In addition, one can use Eq.(\ref{CBL}) to analyze SET experiments. For example, B. Dutta et al. \cite{Dutta2017} reported the measurements of heat and charge transport through the symmetric SETs called samples A and B, with the parallel dimensionless conductance, $ g = 0.629 $ (low-conductance SET) and $ g = 1.98 $ (high-conductance SET), respectively. Eq.(\ref{CBL}) shows that the samples A and B are in Coulomb blockade regions for operating with $\beta E_{C} > 2.187$  and $\beta E_{C} > 2.029$, respectively. However, they operated the SETs with $\beta E_{C} > 21.87$ and $\beta E_{C} > 20.19$, corresponding with $132 \,\,mK$ and $152 \,\, mK$, where $E_{C}=155 \,\mu eV$ and $E_{C}= 100 \,\mu eV$, respectively. That means they operated at about ten times the boundary temperature for their investigations. As described above, the CBBL, determined by the essential parameters  $ g $ and $\beta E_{C} $, holds significant relevance for both experimental fabrications and theoretical investigations of SETs.

\section{Conclusions}
In this paper, we have proposed a method to calculate the CBPD of a SET by utilizing the DC Monte Carlo data. Using the PIMC technique, we have demonstrated that the parallel dimensionless conductance and temperature influence the SET's DC expressed in terms of the expected values of the winding number. Furthermore, our study established the range of minimum and maximum DC, providing a means to quantify the strength of the Coulomb blockade effect through the visibility parameter. For example, the CBBL was calculated with a visibility parameter of $\mathbf{V}=10\%$. The resulting boundary line effectively delineates the transition between the Coulomb and non-Coulomb blockade regimes, allowing for the establishment of a comprehensive CBPD. In addition, SET experiments were introduced to verify the CBPD. We found that the equation of the CBBL can predict the Coulomb blockade effect occurring in both the low and high-conductance SET. Further, the method presented in this study hold the potential to serve as a standard framework for investigating SET and quantum island systems with increasingly complex topologies.

\section*{Acknowledgments}
\label{sec:acknow}
This research project was financially supported by Mahasarakham University for the main funding and the NanoMaterials Physics Research Unit (NMPRU). We especially thank K. Limtragool for valuable discussions.

\section*{References}

\bibliographystyle{iopart-num}
\bibliography{REF}

\end{document}